\documentstyle[11pt]{article}
\begin{document}
\def\dx{\partial_x}
\def\deltamn{\delta_{m+n,0}}
\def\deltaxy{\delta(x-y)}
\def\levi{\epsilon_{ijk}}

\def\rlx{\relax\leavevmode}
\def\inbar{\vrule height1.5ex width.4pt depth0pt}
\def\IZ{\rlx\hbox{\small \sf Z\kern-.4em Z}}
\def\IR{\rlx\hbox{\rm I\kern-.18em R}}
\def\ID{\rlx\hbox{\rm I\kern-.18em D}}
\def\IC{\rlx\hbox{\,$\inbar\kern-.3em{\rm C}$}}
\def\IN{\rlx\hbox{\rm I\kern-.18em N}}
\def\one{\hbox{{1}\kern-.25em\hbox{l}}}
\def\smallfrac#1#2{\mbox{\small $\frac{#1}{#2}$}}

\begin{titlepage}

May, 1995 \hfill{UTAS-PHYS-95-17}\\
\mbox{}\hfill{hep-th/9506129}
\vskip 1.6in
\begin{center}
{\Large {\bf Covariant scalar representation of $iosp(d,2/2)$  }}\\[5pt]
{\Large  {\bf and }}\\[5pt]
{\Large {\bf quantization of the
scalar relativistic particle}}
\end{center}

\normalsize
\vskip .4in

\begin{center}
P. D. Jarvis  \hspace{3pt}
and \hspace{3pt} I Tsohantjis
\par \vskip .1in \noindent
{\it Department of Physics, University of Tasmania}\\
{\it GPO Box 252C Hobart, Australia 7001}
\end{center}
\par \vskip .3in

\begin{center}
{\Large {\bf Abstract}}\\
\end{center}

\vspace{1cm}

A covariant scalar representation of $iosp(d,2/2)$ is constructed and
analysed in comparison with existing methods for the quantization of the
scalar relativistic particle. It is found that, with appropriately defined
wavefunctions, this $iosp(d,2/2)$ produced representation
can be identified with the state space arising from the canonical
BFV-BRST quantization of the
modular invariant, unoriented scalar particle (or antiparticle) with
admissible gauge fixing conditions. For this model, the cohomological
determination of
physical states can thus be obtained purely from the representation theory of
the $iosp(d,2/2)$ algebra.

\end{titlepage}

\section{Introduction and Main Results}

The understanding of the quantization problem for systems with constraints
has had a long development since the seminal monographs of Dirac\cite{Dirac}.
The techniques introduced to handle gauge theories such as
nonabelian Yang-Mills-Shaw theory and (linearised) gravity
culminated in the demonstration of global supersymmetries\cite{BRST}
for such systems, under which gauge and ghost degrees of freedom
transform, and which also play a role even
at the level of classical dynamics with finitely many degrees of freedom.
In certain cases it is possible to unify further these `quantisation'
supersymmetries with other
symmetries possessed by the system, particularly those associated with the
constraint algebra, so that the entire state space may be constructed
from the representation theory of the enlarged algebra (see below). The
ultimate goal of
such work is that sufficient understanding of the gauge symmetries themselves,
the nature of their graded extensions, and the associated representation
theory,
may enable admissible quantisation(s) to be implemented systematically (and
covariantly)
at this algebraic level.

In the present paper,
some preliminary steps in this direction are taken: the attitude adopted is
that
the general principles of this algebraic version of the
quantisation programme should emerge from detailed consideration of particular
case
studies. The initial example taken up below, is a quantum mechanical one, that
of the scalar
relativistic particle. In a following paper\cite{pdjit}, it is intended to
extend the analysis to the spinning particle. The enlarged
algebra in these cases turns out to be an orthosymplectic extension of
the Poincar\'{e} space-time symmetry algebra. Subsequent papers in this series
will consider other first quantised models,
as well as second-quantised gauge field theories, for which the full structure
of the
extended algebra is not yet established.

Before proceeding to discuss the details of the paper and the main results,
it is useful to give a brief historical review of the evolution of
understanding of
the nature of extended symmetries for constraint quantisation. Following
the introduction of scalar-vector spacetime supersymmetries in field theory in
connection with
critical systems\cite{parisi} and with gauged internal
superalgebras\cite{dondijarvis}
the first presentations of BRST\cite{BRST} and anti-BRST\cite{ojima}
transformations in
superspace\cite{bonora} were given a covariant $osp(d\!-\!1,1/2)$ formulation
for
Yang-Mills-Shaw theory and gravity\cite{jarvisetal}, in which the ghost fields
were leading terms in superfield expansions of the graded
components of the `superpotential', and the BRST operators are
supertranslations.
Such formulations have recently been used in discussions of
renormalization and Ward identities\cite{joglekar}, and in discussions of
higher derivative
field theories\cite{ishizuka}.

With the development of the BFV approach\cite{BFV} to canonical quantisation
of systems with open gauge algebras arises the issue
of extended quantisation symmetries also in this context. For the scalar
relativistic
particle, following earlier analysis\cite{monaghan} on the compatibility of
boundary conditions and gauge fixing terms, it was
shown\cite{neveuwest,casalbuoni1}
that the action following from the BFV-BRST canonical analysis does indeed
possess
an extended spacetime supersymmetry, with respect to $iosp(d,2/2)$; this was
extended to the first quantisation of the spinning particle, the galilean
particle and the
massless conformally invariant particle\cite{casalbuoni2}
and also to the bosonic string\cite{neveuwest}.
More general approaches to covariant quantisation and string field theory
involving orthosymplectic spacetime supersymmetries have also been
given\cite{siegel, kallosh, spence, superstring}. Algebraic aspects of
the BFV-BRST extended constraint algebra have been discussed in general,
leading to the expectation that\cite{niemi} $osp(1,1/2)$ or\cite{vanholten}
$igl(1/1)$
symmetries are always realised; the bosonic string would then be
expected\cite{niemi}
to possess a quantisation covariant with respect to $osp(26,2/2)$.

In the present paper our aim is to give a detailed analysis of the
extended $iosp(d,2/2)$ spacetime quantisation symmetry of the
relativistic scalar particle in $d$ dimensional Minkowski space. In recent work
Cornwell and Hartley\cite{hartley,hartleyJFC} have developed formal aspects of
the representation
theory of orthosymplectic superalgebras, and this forms the basis of our
construction.
Specifically, we develop (\S 2 below) a certain massless (irreducible)
covariant scalar produced algebra module. This is then compared (\S 3)
with the state space arising from the quantisation of the scalar relativistic
particle,
following the detailed analyses of Govaerts\cite{govaertsbk}.
After appropriate canonical transformations of variables, and identification of
wavefunctions, the
respective algebra actions are shown to be homomorphic.
Concluding remarks and outlook for further work are given in \S 4 below.

The major result of our analysis is thus that the quantisation and
(cohomological) identification
of physical sates can be obtained for this model, purely from the
representation theory of
the $iosp(d,2/2)$ algebra. In concluding this introduction, it should be
pointed out that our
approach does not require a superfield formalism
(Grassmann variables arise only as dynamical degrees of freedom at the
classical level
in the BFV method), the produced algebra representations being developed
explicitly
in terms of appropriate multiplets of wavefunctions.
Further, the $iosp(d,2/2)$ covariance is shown directly for the state space,
rather than via the derived phase or configuration space path integral
representations,
as has been shown in other approaches\cite{casalbuoni1, neveuwest}.
In fact, issues of gauge invariance for physical states and their inner
products
certainly arise at the canonical level. As will be discussed further below,
their resolution requires taking explicit account of Teichm\"{u}ller space
and modular invariance for this problem.  The module homomorphism
is between (one of two types of) produced $iosp(d,2/2)$ representation, and,
in technical terms\cite{govaertsbk}, the BFV-BRST
canonical quantisation of the modular invariant fundamental hamiltonian
description
of the unoriented scalar relativistic particle (or antiparticle, respectively).

\section{Representation theory of $iosp(d,2/2)$}

In this section we discuss those elements of the representation
theory of inhomo-geneous super-algebras\cite{hartley,hartleyJFC} which will be
needed for our
algebraic consruction of the superparticle quantisation using the superalgebra
$iosp(d,2/2)$.
The abstract theory of induced representations for this case will be treated
in a separate work.

\subsection*{Notation}

The $iosp(d,2/2)$ superalgebra is a generalization of $iso(d,2)$.
The metric tensor $g$ of $iosp(d,2/2)$ has a diagonal block form with the
entries
being the metric tensor of $so(d,2)$ with -1 occurring $d\!-\!1$ times,
$g_{ab}= $ ${\rm diag}(1,-1,\ldots,-1,1)$,
and the symplectic metric tensor being given by
$\epsilon_{12}=-\epsilon_{21}=i$
and $\epsilon^{{\alpha}{\beta}}=\epsilon_{{\alpha}{\beta}}$.
Here latin indices take values $0,1, \ldots, d-1,d,d+1$,
unless otherwise specified, and greek indices $\alpha, \beta, \ldots$ take
values 1,2, while
$\lambda, \mu, \nu \ldots$ take values $0, 1, \ldots, d\!-\!1$
The homogeneous even subalgebra is $so(d,2){\oplus}sp(2,\IR)$. $so(d,2)$ is
generated
by $J_{ab}=-J_{ba}$, and sp(2,\IR) is generated by
$K_{{\alpha}{\beta}}=K_{{\beta}{\alpha}} $. The odd
generators will be denoted by $L_{a\alpha}$. The inhomogeneous part $i(d,2/2)$
consists of $d+2$ even translations $P_{a}$ in the $(d,2)$ pseudoEuclidean
space, and two odd nilpotent translations $Q_{\alpha}$.
The generators can also be expressed  in a light cone basis where we choose,
for the coordinates, momenta and generators
\begin{eqnarray}
x_{\pm} &=& (1/(\sqrt{2})(x_{d+1} {\pm} x_{d}) \nonumber \\
P_{\pm} &=&(1/(\sqrt{2})(P_{d+1} {\pm}P_{d}) \nonumber \\
J_{{\pm}a} &=& (1/\sqrt{2})(J_{(d+1)a} {\pm} J_{(d)a}) \nonumber \\
L_{{\pm}{\alpha}} &=& (1/\sqrt{2})(L_{(d+1){\alpha}} {\pm} L_{d{\alpha}})
\end{eqnarray}
Such a choice is not accidental as will become apparent
later. In this case latin indices $a,b = 0,1, \ldots, d-1,+,- $, while
$g_{ab} = diag(1,-1,\cdots -1)$ and  $g_{+-}=g_{-+}=1$.
The nonzero $iosp(d,2/2)$ commutation relations in the light cone choice read
as follows \cite{hartley,hartleyJFC}:
\begin{eqnarray}
{[}J_{ab} , J_{cd}{]}&=&-i(g_{ac}J_{bd}-g_{bc}J_{ad}+g_{bd}J_{ac}-g_{ad}J_{bc})
\nonumber \\
{[}K_{\alpha\beta} ,
K_{\gamma\delta}{]}&=&-({\epsilon}_{\alpha\gamma}K_{\beta\delta}
+{\epsilon}_{\beta\gamma}K_{\alpha\delta}+{\epsilon}_{\beta\delta}K_{\alpha\gamma}
+{\epsilon}_{\alpha\delta}K_{\beta\gamma}) \nonumber \\
{[}J_{ab} , L_{c{\alpha}}{]}=-i(g_{ac}L_{b{\alpha}}-g_{bc}L_{a{\alpha}})&&
{[}K_{\alpha\beta} , L_{a{\gamma}}{]}=-({\epsilon}_{\alpha\gamma}
L_{a{\beta}}+{\epsilon}_{\beta\gamma}L_{a{\alpha}}) \nonumber \\
{[}L_{a{\alpha}} , L_{b{\beta}}{]}&=&i({\epsilon}_{\alpha\beta}J_{ab} -ig_{ab}
 K_{\alpha\beta}) \nonumber \\
{[}J_{ab} , P_{c}{]}=-i(g_{ac}P_b - g_{bc}P_a)&&
{[}K_{\alpha\beta} ,Q_{\gamma}{]}=-({\epsilon}_{\alpha\gamma}Q_{\beta} +
{\epsilon}_{\beta\gamma}Q_{\alpha}) \nonumber \\
{[}L_{a{\alpha}} , P_{c}{]}=-ig_{ac}Q_{\alpha}&&
{[}L_{a{\alpha}} , Q_{\beta}{]}=-i{\epsilon}_{\alpha\beta}P_{a}.
\label{iospalg}
\end{eqnarray}
It should be noted that the generators satisfying the above
algebra are those of the complexification of $iosp(d,2/2)$.
That is, they are linearly independent over $\IR$ and $\IC$. Moreover they
can be considered as a basis of $iosp(d+2/2)$, the inhomogeneous
extension of the compact real form of
 an appropriate basic classical simple complex Lie
 superalgebra.\footnote{Writing the basis elements which
are linearly independent over $\IR$, and thus form the real $iosp(d,2/2)$, as
$M_{ab}=iJ_{ab}$,$M_{{\alpha}{\beta}}=iK_{{\alpha}{\beta}}$,
$M_{a{\beta}}=e^{i{\pi}/4}L_{a{\beta}}$,
$R_{a}=i{\zeta}P_a$,
$R_{\alpha}=e^{i{\pi}/4}{\zeta}Q_{\alpha}$,  the commutation relations read
${[}M_{AB} , M_{CD}{]}= C_{AB,CD}^{EF}M_{EF}$, ${[}M_{AB},R_C{]}=
C_{AB,C}^{D}R_{D}$,
$\zeta$ being an arbitrary non zero
real constant, with capital Latin indices in the range $0,1, \ldots, d+4$,
$g_{(\alpha+d+2) (\beta+d+2)} = \epsilon_{\alpha \beta}$, and where the
structure
constants are built covariantly from $g_{AB}$ and $\delta_A^B$
with appropriate symmetry and grading factors. Similarly the Pauli-Loubanski
operator
can be written covariantly as $W_{ABC} = M_{AB}R_C + \cdots$}
$osp(d,2/2)$ is the non-compact
real form of one of the basic classical simple complex Lie
 superalgebras
$B(m,1)$ or $D(m,1)$\cite{corn}. It can be obtained from an appropriate
automorphism
of the compact real forms of the above mentioned superalgebras\cite{park}.
A realization of $osp(d,2/2)$ is provided by the $d+4$ dimensional
supermatices $M$ satisfying
$$M^{st}g - (-1)^{[M]}gM=0,$$
where $g$ is the metric tensor (see footnote) and ${[}M{]}$ is 0 (for even
supermatrices)
or 1 (for odd supermatrices) respectively.

The obvious quadratic Casimir operator (the analogue of the mass operator in
the
Poincar\'{e} case) is
\begin{equation}
C_{2}=P_aP^a + Q_{\alpha}Q^{\alpha}.
\end{equation}
A generalized Pauli-Loubanski operator has been found, and the fourth order
Casimir is given by
\begin{equation}
C_4 = \frac{1}{3}W_{abc}W^{abc} - W_{ab{\alpha}}W^{ab{\alpha}}  +
W_{a{\alpha}{\beta}}W^{a{\alpha}{\beta}} -
\frac{1}{3}W_{{\alpha}{\beta}{\gamma}}W^{{\alpha}{\beta}{\gamma}},
\end{equation}
where
\begin{eqnarray}
W_{abc}&=& J_{ab}P_c +  J_{bc}P_a +J_{ca}P_b \nonumber \\
W_{ab{\alpha}}&=& J_{ab}Q_{\alpha} + L_{a{\alpha}}P_b
-L_{b{\alpha}}P_a \nonumber \\
W_{a{\alpha}{\beta}} &=& iL_{a{\alpha}}Q_{\beta} + K_{{\alpha}{\beta}}P_a
+ iL_{a{\beta}}Q_{\alpha} \nonumber \\
W_{{\alpha}{\beta}{\gamma}} &=& K_{{\alpha}{\beta}}Q_{\gamma} +
K_{{\beta}{\gamma}}Q_{\alpha}+K_{{\gamma\alpha}}Q_{\beta}.
\end{eqnarray}

\subsection*{The covariant scalar multiplet.}

We now turn to the construction of the covariant scalar multiplet,
adapting the exposition of Hartley and Cornwell\cite{hartley,hartleyJFC}.
Let us start with the definition the
covariant representations of the group $ISO(d,2)$ which follows exactly the
same lines of exposition as that of the normal Poincar\'{e} group.
It should also be noted that, although not directly used, we should deal
with the universal covering group of proper orthochronous $ISO_0(d,2)$.
The $d\!+\!2$ dimensional pseudo-Euclidean space is identified with the coset
space
$ISO(d,2)/SO(d,2)$. We shall denote a general element of $ISO(d,2)$ by
$(t, {\Lambda})$ where $(0, {\Lambda})$ is a rotation and $(t,1)$ a translation
on the space. The identity, inverse, product, and the action of
$ISO(d,2)$ on the manifold are respectively given by $(0,1)$,
 $(t, {\Lambda})^{-1} = (-{\Lambda}^{-1}t,{\Lambda}^{-1})$,
$(t, {\Lambda})(t', {\Lambda}')=(t+{\Lambda}t', {\Lambda}{\Lambda}')$ and
$(t, {\Lambda})x ={\Lambda}x+t$.
Let ${\Gamma}'_{0}$ be a finite dimensional representation
of $SO(d,2)$ carried by infinitely differentiable Borel functions
${\phi}(x)$ for any point $x \equiv (x^a) \equiv (x^{\mu},x^{d},x^{d+1})$ , and
taking
values in $\IC$. We shall denote the carrier space by
$V'_{0} = \IC^{\infty}(ISO(d,2)/SO(d,2),\IC)$. ${\Phi}'_{0}(t, {\Lambda})$ will
denote the operators of the representation corresponding to an element
$(t, {\Lambda})$ of $ISO(d,2)$, and the representation will be denoted by the
pair $({\Phi}'_{0}, V'_{0})$.
The covariant representation of $ISO(d,2)$ is a representation induced from
the representation ${\Gamma}'_{0}$ of $SO(d,2)$ given by
\begin{equation}
{\Phi}'_{0}(t,
{\Lambda}){\phi}'_{0}(x)={\Gamma}'_{0}({\Lambda}){\phi}'_{0}({\Lambda}^{-1}(x-t)).
\label{inducedrep}
\end{equation}
In the case of a scalar representation ${\Gamma}'_{0}({\Lambda})=I$.
This representation provides as usual a representation of the
algebra $iso(d,2)$ given by
\begin{eqnarray}
{\Phi}'_{0}(J_{ab}){\phi}'_{0}(x)&=&i(x_{a}\frac{\partial}{\partial x^{b}}-
x_{b}\frac{\partial}{\partial x^{a}}){\phi}'_{0}(x) +
{\Gamma}'_{0}(J_{ab}){\phi}'_{0}(x),\\
{\Phi}'_{0}(P_{a}){\phi}'_{0}(x)&=&i\frac{\partial}{\partial
x^{a}}{\phi}'_{0}(x)
\end{eqnarray}
This representation extends naturally to a representation of the universal
enveloping algebra $U(iso(d,2))$ by
defining ${\Phi}'_{0}(1){\phi}'_{0}(x)={\phi}'_{0}(x)$, 1 being the identity of
$U(iso(d,2))$.
Again for a scalar representation, ${\Gamma}'_{0}(J_{ab}) =0$.

According to \cite{hartley} and \cite{blattner}, the above representation is
equivalent
to a representation of $iso(d,2)$ produced from the representation
${\Gamma}'_{0}$ of its subalgebra $so(d,2)$, defined as follows.  Let
$U(iso(d,2))$ be regarded
as a left $U(so(d,2))$-module. This means
that the basis of $U(iso(d,2))$ will be of the form
\begin{equation}
P^r={\prod}{P_{0}}^{r_{0}}{P_{1}}^{r_{1}}\ldots
{P_{d}}^{r_{d}}{P_{d+1}}^{r_{d+1}}
\end{equation}
for all $r =(r_{0},r_{1}...r_{d}, r_{d+1}) {\in}{\IN^{d+2}}$, and a general
element $X$ of $U(iso(d,2))$ is given by
\begin{equation}
X={\sum} A_r P^r
\label{Umoduledef}
\end{equation}
where $A_r {\in} U(so(d,2))$. ${\Gamma}'_{0}$ is carried by infinitely
differentiable functions defined on $U(iso(d,2))$ regarded as a left
$U(so(d,2))$ module,
and taking values in $\IC$.  We shall denote this space of functions by
$V_{0}=Hom_{U(so(d,2)}( {\cal P}, \IC)$ where ${\cal P}$ is the real vector
space spanned by
all combinations of $P^{r}$.
Then the produced algebra representations are defined for $\phi_{0}{\in}V_{0}$
by
\begin{equation}
\Phi_{0}(X)\phi_{0}(Y)=\phi_{0}(YX) \quad {\mbox{\rm and}}\quad
\phi_0(AX)=\Gamma_{0}'(A)\phi_0(X),
\label{producedrep}
\end{equation}
where $X,Y {\in} U(iso(d,2))$ and $A{\in} U(so(d,2))$.
It can also be shown\cite{hartley} that this definition is equivalent
to the following definition of produced representations:
\begin{equation}
\Phi(X){\phi}_{0}(P)={\sum}{\Gamma}_{0}'((PX)_r){\phi}_{0}(P^r),
\label{altproducedrep}
\end{equation}
where $X{\in}iso(d,2)$, $P\in {\cal P}$ and $(PX)_r{\in}U(so(d,2))$ are to be
interpreted
as the $U(so(d,2))$-combinations of $PX$ in
$U(iso(d,2))$ regarded as an $U(so(d,2))$-module (see (\ref{Umoduledef})).
Following \cite{hartleyJFC,blattner}, for each  element
${\phi}'_{0}{\in}V'_{0}$, we define
a function $\phi_{0}$ by
\begin{equation}
\phi_{0}(X)={\Phi}'_{0}(X){\phi}'_0(x_0)
\label{fnequivalence}
\end{equation}
where $x_0 {\in}ISO(d,2)/SO(d,2)$ is stable under $SO(d,2)$. Then $\phi_{0}$
satisfies the definition (\ref{producedrep}) of the produced algebra
representation, and
lies in $V_{0}$.
Also for a ${\psi}'_{0}$ such that ${\psi}'_{0}=  {\Phi}'_{0}(X){\phi}'_{0}$
for some $X$ of $iso(d,2)$, then there exists a ${\phi}_{0}$ of $V_{0}$ defined
by
(\ref{fnequivalence}) using
${\psi}'_{0}$ above, and satisfying ${\psi}_{0}  = \Phi_{0}(X){\phi}_{0}$.
Thus the representations $({\Phi}'_{0},V'_{0})$ and $({\Phi}_{0},V_{0})$ are
equivalent;
in particular, ${\phi}_0'(x_0 ) = {\phi}_0(1)$.
An explicit realization of functions ${\phi}_0'(x)$ expressed in terms of
${\phi}_{0}(P^{r})$ that exhibits the above equivalence is given by:
$${\phi}_{0}'(x) = {\prod}_a{\sum}_{r_a}
(1/{r_a}!)(-ix^{a})^{r_a}{\phi}_{0}(P^{r})$$
where $a= 0, 1, \ldots, d, d+1$ and $x^{a}$ take any real value.
Then it can be shown, using the definition of
the produced representation (\ref{producedrep}) that relations
(\ref{inducedrep}) are satisfied.

We can now proceed to construct the representation $({\phi}, V)$
of $iosp(d,2/2)$
produced by the trivial representation of $osp(d,2/2)$. This is precisely
what should be called a covariant scalar representation of $iosp(d,2/2)$.
The definition of the produced Lie superalgebra representations is the same
as for Lie algebras mentioned above.
The $U(iosp(d,2/2))$ regarded as a $U(osp(d,2/2))$ module has basis of the form
$P^rQ^s$ with $P^r$ as in (\ref{Umoduledef}) and
$Q^s=Q^{s_{1}}_{1}{Q^{s_{2}}_{2}}$
where $s_{1}$, $s_{2}{\in}(0,1)$ and $s{\in}(0,1){\times}(0,1)$.
Let $\Gamma$ be a representation of $osp(d,2/2)$.
The carrier space consists of linear functions defined on $P^rQ^s$ and
thus $V=Hom_{U(iosp(d,2))}({\cal P'},\IC)$ where ${\cal P'}$ is spanned by
real combinations of the
basis elements $P^rQ^s$. The produced superalgebra representation is defined by
\begin{eqnarray}
\Phi(X){\phi}(P^rQ^s)&=&\phi(P^rQ^sX), \nonumber \\
\phi(AP^rQ^s)&=&{\Gamma}(A){\phi(P^rQ^s)}
\end{eqnarray}
where $A{\in}U(osp(d,2/2))$, $X{\in}iosp(d,2/2)$ and $\phi{\in}V$.
For the covariant scalar representation the $osp(d,2/2)$ is represented
trivially and thus ${\Gamma}$ =0 ($={\Gamma}'$ when restricted to $so(d,2)$).
Note now that every $\phi{\in}V$, when defined on ${\cal P}{\subset}{\cal P'}$,
is a
member of $V_{0}$, and via the equivalence mentioned above
gives a member of $V'_{0}$. Moreover, from the definition of produced
superalgebra representation, there is a one to one equivalence between a
$\phi{\in}V$ and a set of four functions defined solely on $P^r$, namely
$\phi(P^r)$, $(\Phi(Q_{\alpha}){\phi})$,
$(\Phi(Q_{\alpha}Q_{\beta}){\phi})$.
Thus, using this we regard an element of $V$ as comprising the following set of
four functions
defined on $ISO(d,2)/SO(d,2)$:
\begin{equation}
\phi(x),\quad (\Phi(Q_{\alpha}){\phi})(x)={\phi}(x,\alpha),\quad
(\Phi(Q_{1}Q_{2}){\phi})(x)={\phi}(x,12).
\end{equation}
Finally, the action of the operators $\Phi(X)$ for every
$X{\in}iosp(d,2/2)$
can be evaluated by calculating $\Phi(X)$ on these four functions, using the
relations (14), (11), (7-8), the above realization of ${\phi}_{0}(x)$
and the commutation relations
(\ref{iospalg})(the dashes have been dropped out via the equivalence
 mentioned above).
This action for the covariant $iosp(d,2/2)$ scalar multiplet is given by
\begin{eqnarray}
{\Phi}(J_{ab}){\phi}(x)={\Phi}_0(J_{ab}){\phi}(x)&&
{\Phi}(P_{a}){\phi}(x)={\Phi}_0(P_{a}){\phi}(x) \nonumber \\
{\Phi}(J_{ab}){\phi}(x,{\alpha})={\Phi}_0(J_{ab}){\phi}(x,{\alpha})&&
{\Phi}(P_{a}){\phi}(x,{\alpha})={\Phi}_0(P_{a}){\phi}(x,{\alpha})
\quad {\alpha} = 1,2 \nonumber \\
{\Phi}(J_{ab}){\phi}(x,{\alpha\beta})={\Phi}_0(J_{ab}){\phi}(x,{\alpha\beta})&&
{\Phi}(P_{a}){\phi}(x,{\alpha\beta})={\Phi}_0(P_{a}){\phi}(x,{\alpha\beta})
\quad {\alpha}, {\beta}=1,2 \nonumber \\
{\Phi}(K_{\alpha\beta}){\phi}(x)=0&&
{\Phi}(Q_{\alpha}){\phi}(x)={\phi}(x,{\alpha}) \nonumber \\
{\Phi}(K_{\alpha\beta}){\phi}(x,{\gamma})={\epsilon}_{\alpha\gamma}
{\phi}(x,{\beta})+ {\epsilon}_{\beta\gamma}{\phi}(x,{\alpha})&&
{\Phi}(Q_{\alpha}){\phi}(x,{\beta})=-{\phi}(x,{\alpha\beta}) \nonumber \\
{\Phi}(K_{\alpha\beta}){\phi}(x,{\beta\gamma})=0&&
{\Phi}(Q_{\alpha}){\phi}(x,{\beta\gamma})=0 \nonumber \\
{\Phi}(L_{a{\alpha}}){\phi}(x)&=&g_{ab}x^{b}{\phi}(x,{\alpha}) \nonumber \\
{\Phi}(L_{a{\alpha}}){\phi}(x,{\beta})&=&-g_{ab}x^{b}{\phi}(x,{\alpha\beta})
-i{\epsilon}_{\alpha\beta}{\Phi}_{0}(P_{a}){\phi}(x) \nonumber \\
{\Phi}(L_{a{\alpha}}){\phi}(x,{\beta\gamma})
&=&-i{\epsilon}_{\beta\gamma}{\Phi}_{0}(P_{a}){\phi}(x,{\alpha})
\end{eqnarray}

An indefinite inner product is given by \cite{hartley,hartleyJFC}
\begin{equation}
({\phi} , {\psi}) =\int
d^{d\!+\!2}x{\Omega}^{{\alpha\beta}}{[}{\phi}^{*}(x,{\alpha\beta})
{\psi}(x) - {\phi}^{*}(x){\psi}(x,{\alpha\beta})
-{\phi}^{*}(x,{\alpha}){\psi}(x,{\beta})+
{\phi}^{*}(x,{\beta}){\psi}(x,{\alpha}){]}
\label{prodinprod}
\end{equation}
Under this inner product, for functions with appropriate boundary conditions,
the $iso(d,2)$ and $sp(2,\IR)$ generators are represented by
Hermitian operators while the rest are antiHermitian.

Irreducibility of the covariant scalar multiplet demands that each of the
Casimir operators has the same eigenvalue on all the four functions
above. In particular we demand that
\begin{eqnarray}
C_{2}{\phi}(x)&=&{\Phi}(P_aP^a){\phi}(x) +
{\Phi}(Q_{\alpha}Q^{\alpha}){\phi}(x) \nonumber \\
&=&{\Phi}(P_aP^a){\phi}(x) + 2i{\phi}(x,12) = {\lambda}{\phi}(x), \nonumber \\
C_{2}{\phi}(x,{\alpha})&=&{\Phi}(P_aP^a){\phi}(x,{\alpha}) = {\lambda}
{\phi}(x,{\alpha}), \nonumber \\
C_{2}{\phi}(x,{\alpha\beta})&=&{\Phi}(P_a P^a){\phi}(x,{\alpha\beta})=
{\lambda}{\phi}(x,{\alpha\beta}).
\label{Caseigenvalue}
\end{eqnarray}
where $\lambda$ is the constant eigenvalue of $C_{2}$ characterising the
irreducible $iosp(d,2/2)$ multiplet.
Finally we can introduce an (antiHermitian) ghost number operator given by
\begin{equation}
Q_{c}=i/2(K_{11} -K_{22})
\end{equation}
so that the functions of definite ghost number are: ${\phi}(x)$ and
${\phi}(x,12)$
with ghost number 0, and
${\phi}(x,1) \pm {\phi}(x,2)$ with $\mp 1$, respectively.

\section{BFV-BRST quantisation of the scalar relativistic particle}

As is well known\cite{govaertsbk,henneaux} the BFV canonical quantisation of
constrained
Hamiltonian systems\cite{BFV} uses an extended phase space description in
which, to each first class constraint, a pair of conjugate `ghost' variables
(of Grassmann parity opposite to that of the constraint) is introduced.
Here we follow this procedure for the scalar relativistic particle. Although
our notation
is adapted to the massive case, $m >0$, as would follow from the second
order action corresponding to extremisation of the proper length of the
particle world line,
an analysis of the {\it fundamental} Hamiltonian description of the
first order action\cite{govaertsbk} leads to an equivalent picture
(with an additional mass parameter $\mu \ne 0$ supplanting $m$
in appropriate equations, and permitting $m\rightarrow 0$ as a smooth limit).
In either case, for the scalar particle the primary first class constraint is
the mass-shell condition $(P^2 - m^2)$, where $P^2 = P_{\mu}P^{\mu}$; including
the corresponding
Lagrange multiplier $\lambda$ as an additional dynamical variable then leads to
a secondary constraint, reflecting conservation of its conjugate momentum.
The quantum formulation, to which we proceed directly,
should be consistent with the equations of motion
and gauge fixing at the classical level. We choose below to work in the class
\cite{monaghan,teitelboim} $\dot{\lambda}=0$; moreover, with the restriction to
{\it orientation preserving} diffeomorphisms (world-line reparametrizations),
it is sufficient to choose $\lambda > 0$ (a parallel treatment applies for
$\lambda < 0$).
This restriction will also be essential in establishing the equivalence to the
algebraic
approach of the \S 2 above.

\subsection*{State space and wavefunctions.}

The BFV extended phase space\cite{govaertsbk} for the BRST quantisation
of the scalar relativistic particle is taken to comprise
the following canonical variables:
\begin{equation}
x^{\mu}({\tau}), p_{\mu}({\tau}), \quad {\omega}({\tau}), {\pi}({\tau}),
\quad {\eta}^{\alpha}({\tau}), {\rho}_{\alpha}({\tau})
\end{equation}
where $\omega$ parametrizes
the Lagrange multiplier $\lambda = e^\omega$,
$\pi$ is the momentum conjugate to $\omega$, and ${\eta}^{\alpha}$,
${\rho}_{\alpha}$,
${\alpha}=1,2$ are the Grassmann odd BFV extended phase space variables.
The operators corresponding to the above set satisfy the following commutation
relations\footnote{The choice of $\omega$ and $\pi$ as conjugate variables
corresponds to a choice of a particular inner product, and hence Hermitian
canonical
conjugate to $\lambda$, for the direct problem of quantisation
on the half-line $\lambda > 0$. The ultimate determinant of these choices
is the identification with the produced representation of \S 2 above (see
below).}:
\begin{eqnarray}
{[}X_{\mu}, P_{\nu}{]} &=& -ig_{\mu \nu} \nonumber \\
{[}{\hat{\omega}}, {\hat{\pi}}{]} &=& i  \nonumber \\
{[}{\eta}^{\alpha}, {\rho}_{\beta}{]} &=& -i{\delta}^{\alpha}_{\beta}
\label{xpalg}
\end{eqnarray}
The ghost number operator $Q_{c}$  is defined by
\begin{equation}
Q_{c}=(i/2)({\eta}^{\alpha}{\rho}_{\alpha}-{\rho}_{\alpha}{\eta}^{\alpha}).
\label{Qc}
\end{equation}
The canonical BRST operator is given by
\begin{equation}
{\Omega} = {\eta}^{1}{\hat{\pi}} + {\eta}^{2}(P^2 - m^2);
\end{equation}
we shall also use the corresponding anti-BRST operator
\begin{equation}
\bar{\Omega} = {\frac{i}{2}}({\rho}_{2}{\hat{\pi}} - {\rho}_{1}(P^2 - m^2)).
\end{equation}
The gauge fixing operator\cite{BFV} ${\Psi}$ which will lead to the appropriate
effective Hamiltonian is given by:
\begin{eqnarray}
{\Psi} &=& -{\frac{1}{2}}e^{\hat{\omega}}{\rho}_{2},  \quad  \mbox{{\rm and}}
\nonumber \\
H &=& i[{\Psi} , {\Omega}] =
-{\frac{1}{2}}(e^{{\hat{\omega}}}{\eta}^{1}{\rho}_{2} +
e^{{\hat{\omega}}}(P^2 - m^2)).
\end{eqnarray}

Consider the linear representation of the algebra of (\ref{xpalg}$a$),
(\ref{xpalg}$b$)
on coordinate and momentum space:
\begin{eqnarray}
X^{\mu}|x^{\mu}>  = x^{\mu}|x^{\mu}> &,&  P_{\mu}|x^{\mu}>
      = -i\frac{\partial}{\partial x^{\mu}}|x^{\mu}>,\nonumber \\
P_{\mu}|p_{\mu}> = p_{\mu}|p_{\mu}> &,& X^{\mu}|p_{\mu}>
         = i\frac{\partial}{\partial p_{\mu}}|p_{\mu}>, \nonumber \\
<x^{\mu}{}'|x^{\mu}>  = \delta(x^{\mu}{}'-x^{\mu}) &,&
                    <p_{\mu}{}'|p_{\mu}> = \delta(p_{\mu}'-p_{\mu}), \nonumber
\\
<x^{\mu}|p_{\mu}>  &=& {\frac{1}{(2{\pi})^{d/2}}} e^{-ix^{\mu}p_{\mu}};
\nonumber \\
{\hat{\omega}}|\omega>  = \omega|\omega> &,&
              {\hat{\pi}}|\omega> = i\frac{\partial}{\partial{\omega}}|\omega>,
\nonumber \\
{\hat{\pi}}|\pi> = \pi|\pi> &,&
          {\hat{\omega}}|\pi> = -i\frac{\partial}{\partial{\pi}}|\pi>,
\nonumber \\
<\omega'|\omega>  = \delta(\omega'-\omega) &,&
                      <\pi'|\pi>  = \delta(\pi'-\pi), \nonumber \\
<\omega|\pi>  &=& {\frac{1}{(2{\pi})^{1/2}}}  e^{i{\omega}{\pi}}.
\end{eqnarray}
We also recognise (\ref{xpalg}$c$) as a $b,c$ algebra \cite{govaertsbk},
where $b$ stands for $i{\rho}_{\alpha}$ and $c$ for ${\eta}^{\alpha}$.
Then the algebra admits a
representation on a four dimensional linear space with basis denoted by
$|\pm \pm>, |\pm \mp>$, and the action of ${\eta}^{\alpha}$
and  $i{\rho}_{\alpha}$, is given by
\begin{eqnarray}
{\eta}^{1} |--> = |+->, &\quad & {\eta}^{1} |-+> =|++>  \nonumber \\
{\eta}^{2} |--> = |-+>, &\quad & {\eta}^{2} |+-> =-|++> \nonumber \\
i{\rho}_{1}|+->=|-->,  &\quad & i{\rho}_{1}|++>=|-+> \nonumber \\
i{\rho}_{2} |-+> = |-->, &\quad &  i{\rho}_{2}|++> =-|+->
\end{eqnarray}
The non-zero inner products between these states are given by:
\begin{equation}
<--|++>=-<++|-->=i,\quad <-+|+->=-<+-|-+>=-i
\end{equation}
The above representations and inner products imply the Hermiticity conditions
\begin{eqnarray}
X_{\mu}^{\dag} =X_{\mu} ,\quad P_{\mu}^{\dag}=P_{\mu} ,\quad
{\hat{\omega}}^{\dag} = \omega, \quad   {{\hat{\pi}}}^{\dag}={\pi},\\
({\eta}^{\alpha})^{\dag} = {\eta}^{\alpha}, \quad
({\rho}_{\alpha})^{\dag} = -({\rho}_{\alpha}).
\end{eqnarray}
Finally, with the identity operator given by
\begin{equation}
I= {\sum}_{{\sigma}{\sigma}'={\pm}}{\int}_{\infty} d^dx d{\omega}
(-i)(-1)^{(1-{\sigma}')/2}|x^{\mu},{\omega}, {\sigma},
{\sigma'}><x^{\mu},{\omega}, {-\sigma},{-\sigma'}|
\end{equation}
a general state $|\psi>$ of the system is
\begin{eqnarray}
|\psi> &=& {\sum}_{{\sigma}{\sigma}'={\pm}}{\int}_{\infty} d^dx d{\omega}
|x^{\mu},{\omega}, {\sigma}, {\sigma'}>
{\psi}_{{\sigma}{\sigma'}}(x^{\mu},{\omega},{\tau}), \quad  \mbox{{\rm
where}}  \nonumber \\
{\psi}_{{\sigma}{\sigma'}}(x^{\mu},{\omega},{\tau}) &=&
-i(-1)^{(1-{\sigma}')/2}
<x^{a}, {\omega}, -{\sigma}, -{\sigma'}|\psi>,
\end{eqnarray}
The inner product $<\phi|\psi>$ in terms of wavefunctions is given by
\begin{equation}
<\phi|\psi> = (-i){\int}_{\infty} d^dx d{\omega} (-1)^{(1-{\sigma}')/2}
{\sum}_{{\sigma}{\sigma}'={\pm}}{\phi*}_{{\sigma}{\sigma}'}(x^{\mu},{\omega},{\tau})
{\psi}_{{-\sigma}{-\sigma}'}(x^{\mu},{\omega},{\tau}).
\end{equation}
As usual the wave functions ${\psi}$ are required to vanish at
${\omega} = {\pm \infty}$. With respect to the previously defined ghost number
operator,
(\ref{Qc}),
the kets $|\pm \pm>$, $|\pm \mp>$,
and corresponding wavefunction components $\psi_{\pm \pm}$, $\psi_{\pm \mp}$,
have eigenvalues $\pm 1, 0$ respectively.

The gauge invariant physical states can now be identified\cite{govaertsbk} by
imposing the Schr\"{o}dinger equation
 $i\frac{d}{d{\tau}}|\psi> = H|\psi> \equiv i[{\Psi},{\Omega}]|\psi>$
and computing the cohomology of the
BRST operator $\Omega$. However, in order
to exhibit the $iosp(d,2/2)$ symmetry in the above quantization
procedure at the level of state space, it is convenient to use
equivalent BRST and gauge fixing operators $\Omega'$, $\Psi'$, which can be
more directly
expressed in terms of the superalgebra generators. The wavefunctions
${\psi}_{{\sigma}{\sigma'}}(x^{\mu},{\omega},{\tau})$
can then be readily identified
with those of the functions of \S 2 above which carry the
$iosp(d,2/2)$ produced representation
(with appropriate boundary conditions). Our final identification of physical
states
will then follow with respect to the cohomology of the transformed BRST
operator.

Consider the following canonical transformation on the classical
dynamical variables of the extended phase space\cite{casalbuoni1} :
\begin{eqnarray}
  i\rho'_{\alpha} &=& e^{-\omega}i\rho_{\alpha},  \nonumber \\
{\eta'}^{\alpha} &=& e^{\omega}{\eta}^{\alpha}, \nonumber \\
{\hat{\pi'}} &=& {\hat{\pi}} -({\eta }^{2}\rho_{2} -\rho_{1}{\eta}^{1}) ,
\end{eqnarray}
with the remainder invariant.
At the quantum level the corresponding BRST and anti-BRST operators
${\Omega'} = {\eta'}^{1}{\hat{\pi'}} + {\eta'}^{2}(P^2 - m^2)$,
$\bar{\Omega}' = {\frac{i}{2}}({\rho'}_{2}{\hat{\pi'}} - {\rho'}_{1}(P^2 -
m^2))$
can be written as
\begin{eqnarray}
{\Omega'} = {\eta}^{1}:e^{{\hat{\omega}}}{{\hat{\pi}}}:+
e^{{\hat{\omega}}}{\eta}^{2}(P^2 - m^2)
-e^{{\hat{\omega}}}{\eta}^{2}\rho_{2}{\eta}^{1}, \nonumber \\
{ \bar{\Omega'}} =(i/2)(
\rho_{2}:e^{{\hat{\omega}}}{{\hat{\pi}}}:-e^{{\hat{\omega}}}\rho_{1}(P^2 - m^2)
+e^{{\hat{\omega}}}\rho_{1}\rho_{2}{\eta}^{1}),
\end{eqnarray}
where the symmetric ordering
\begin{eqnarray}
:e^{{\hat{\omega}}}{{\hat{\pi}}}:=(1/2)(e^{{\hat{\omega}}}{{\hat{\pi}}}+
{{\hat{\pi}}}e^{{\hat{\omega}}}) = {{\hat{\pi}}}e^{{\hat{\omega}}} +
(i/2)e^{{\hat{\omega}}}
\end{eqnarray}
has been introduced.
It is also convenient to define\cite{casalbuoni1} $X_{\alpha}$ and
 $Q_{\alpha}$ ($\alpha=1,2$)by
\begin{eqnarray}
Q_{1}=(i/2{\sqrt{2}})(2{\eta}^{1} + i\rho_{2}) &,&
Q_{2}=(i/2{\sqrt{2}})(2{\eta}^{1} - i\rho_{2}) \nonumber \\
X_{1}=(i/\sqrt{2})(i\rho_{1} - 2{\eta}^{2}) &,&
X_{2}=(i/\sqrt{2})(-i\rho_{1} - 2{\eta}^{2}) \nonumber  \\
\mbox{{\rm where}} \quad
{[}Q_{\alpha} , X_{\beta} {]} &=& -i{\epsilon}_{{\alpha}{\beta}}.
\end{eqnarray}

In terms of these variables we attain the following simple forms
for the BRST, gauge fixing and Hamiltonian operators:
\begin{eqnarray}
{\Omega'}&=& (-i/ \sqrt{2})(:\hat{\pi} e^{\hat{\omega}}:(Q_{1}+Q_{2}) +
(X_{1}+X_{2})H), \nonumber \\
{ \bar{\Omega'}}&=&
(-i/{\sqrt{2}})(:\hat{\pi}e^{{\hat{\omega}}}:(Q_{1}-Q_{2}) + (X_{1}-X_{2})H ),
\nonumber \\
{\Psi}' &=& -(1/2)\rho_{2} =(1/ \sqrt{2})(Q_{1}- Q_{2}), \nonumber \\
H'&=&i{[}{\Psi'} , {\Omega}'{]} = -(1/2)e^{{\hat{\omega}}}((P^2 - m^2) +
2iQ_{1}Q_{2}) \equiv H.
\label{newforms}
\end{eqnarray}

\subsection*{Realisation of $iosp(d,2/2)$ superalgebra.}

The realization of $iosp(d,2/2)$ provided by the extended BFV-BRST quantisation
as described
above is formulated in terms of the operators $X^{\mu}$, $P_{\mu}$
together with $Q_{\alpha}$, $X_{\alpha}$ and $X_{+} = {\tau}I$, $P_{-}=H$,
$P_{+}=e^{-{\hat{\omega}}}$,
$X_{-}=:{{\hat{\pi}}}e^{{\hat{\omega}}}:$. Given the commutation relations
\begin{eqnarray*}
&{[}X_{\mu},P_{\nu}{]} = -ig_{{\mu}{\nu}}, \quad {[}X_{-},P_{+}] = i,& \\
&{[}X_{\mu},X_{\pm}{]} ={[}X_{+},P_{-}{]} = {[}X_{+},P_{{\mu}}{]} =
{[}X_{+},P_{+}{]}=
{[}X_{{\mu}},P_{+}{]} = 0,& \\
&{[}Q_{\alpha},X_{\pm}{]} ={[}X_{\alpha},X_{\pm}{]}=
{[}X_{\alpha},P_{+}{]} ={[}Q_{\alpha},P_{\pm}{]} = 0,& \\
&{[}X_{-},P_{-}{]}= -iP^{-1}_{+}P_{-}, \quad
{[}X_{\alpha},P_{-}{]} = iP^{-1}_{+}Q_{\alpha},
\quad {[}X_{{\mu}},P_{-}{]} = iP^{-1}_{+}P_{{\mu}},&
\end{eqnarray*}
it can be checked that the following generators
do indeed satisfy the commutation relations of $osp(d,2/2)$
\begin{eqnarray}
J_{\mu\nu} = X_{\mu}P_{\nu} - X_{\nu}P_{\mu}, \quad
&J_{+-} = X_{-}P_{+}+ X_{+}P_{-},& \quad
J_{\mp \mu} = \mp X_{\mp}P_{\mu}-X_{\mu}P_{\mp},\nonumber \\
K_{{\alpha}{\beta}} = -i(X_{\alpha}Q_{\beta} +
X_{\beta}Q_{\alpha}), \quad
&L_{\mu{\alpha}} = X_{\mu}Q_{\alpha} - X_{\alpha}P_{\mu},& \quad
L_{\mp \alpha} = \mp X_{\mp}Q_{\alpha} - X_{\alpha}P_{\mp},
\label{physalg}
\end{eqnarray}
where $L_{-1}=-i/\sqrt{2}({\Omega'} + { \bar{\Omega'}})$,
$L_{-2}=-i/\sqrt{2}({\Omega'} - { \bar{\Omega'}})$.
Together with $P_\mu$, $P_\pm$, $Q_\alpha$,
these generators
close\footnote{In covariant notation (see footnote to (\ref{iospalg}))
this realization can be written simply in terms of $X_A R_B - (-1)^{|AB|}X_B
R_A$ and $R_A$.
However, as noted above, the $X_A$ and $R_B$ are {\it not}
canonically conjugate.} on the inhomogeneous form $iosp(d,2/2)$ (see
(\ref{iospalg}) above).
It is clear that the $d\!+\!2$-dimensional coordinates $x_\mu$, $x_\pm$,
$x_\alpha$
and momenta $P_\mu$, $P_\mp$, $Q_\alpha$ are {\it not} all canonically
conjugate.
In particular $X_+$, proportional to the identity operator,
simply re-scales kets (at time $\tau$) by $\tau$,
while $P_-$ is identified with the Hamiltonian, a function of the other
variables
(whose action also sets the rate of time development of kets via the
Schr\"{o}dinger equation).

The final stage in the analysis is the identification of the ($\tau$-dependent)
wavefunctions $\psi_{\sigma \sigma'}$ with the functions
over $x^a$ which carry the produced representation in \S 2 above.
To facilitate this comparison we introduce kets and wavefunctions dependent
on $p_+ = e^{-\omega}$ by a change of variables. As $e^{-{\omega}}$ is
a monotonic differentiable
function, $|x^{\mu}, p_+, {\sigma}, {\sigma'}>$ can be defined by
\begin{eqnarray}
|x^{\mu}, {\omega}, {\sigma}, {\sigma'}> &=& p_+^{1/2}
|x^{\mu}, p_+, {\sigma}, {\sigma'}>,
 \quad \mbox{{\rm and where}} \nonumber \\
<p_+|p_+'> &=& {\delta}(p_+ - p_+').
\end{eqnarray}
Then the completeness relation becomes
\begin{eqnarray}
I= {\sum}_{{\sigma}{\sigma}'=
{\pm}}\int d^dx {\int}_{0}^{\infty} dp_{+}
(-i)(-1)^{(1-{\sigma}')/2}|x^{\mu},p_{+}, {\sigma}, {\sigma'}>
<x^{\mu},p_{+}, {-\sigma},{-\sigma'}|
\end{eqnarray}
while
\begin{eqnarray}
{\psi}_{{\sigma}{\sigma'}}(x^{\mu},p_{+},{\tau})&=&
-i(-1)^{(1-{\sigma}')/2}<x^{\mu},p_{+}, -{\sigma}, -
{\sigma'}|\psi(\tau)> \nonumber \\
&=& -i(-1)^{(1-{\sigma}')/2}e^{\omega/2}
{\psi}_{{\sigma}{\sigma'}}(x^{a},{\omega},{\tau})
\end{eqnarray}
It should be noted that the domain of $p_{+}$ is restricted
to be $p_{+}{\in}(0, {\infty})$ as ${\omega}{\in}(-{\infty}, {\infty})$ and
this will result in wave functions
${\psi}_{{\sigma}{\sigma'}}(x^{\mu},p_{+},{\tau})$
which vanish when $p_{+}$ approaches zero or infinity. In the $p_+$
representation
the operator $X_-$ is realized\footnote{The wavefunctions
in the $x_-$ representation are given by
$\psi_{{\sigma}{\sigma'}}(x^{a},x_{-},x_{+}) =
\int dp_{+}e^{-ix_{-}p_{+}}{\psi}_{{\sigma}{\sigma'}}(x^{a},p_{+},x_{+})$.}
as $i\partial/\partial p_+ $, while the inner product becomes
\begin{equation}
<{\phi}|{\psi}>= -i{\int} d^dx {\int}^{\infty}_{0}d{p_{+}}
{\sum}_{{\sigma}{\sigma'}={\pm}}(-1)^{(1+{\sigma}')/2}
{\phi*}_{{\sigma}{\sigma'}}(x^{{\mu}},p_{+},{\tau})
{\psi}_{{-\sigma}{-\sigma'}}(x^{{\mu}},p_{+},{\tau}).
\label{physinprod}
\end{equation}

The action of the operators (\ref{physalg}) on the wavefunctions
${\psi}_{{\sigma}{\sigma'}}(x^{a},p_{+},{\tau})$ is given in the Appendix
together with the Schr\"{o}dinger equation for
them ((see (\ref{XQaction}-\ref{bigschroed})).
It can be easily seen that with the identifications
\begin{eqnarray}
{\phi}(x^{a},p_{+},x_{+})={\psi}_{+-} \nonumber \\
{\phi}(x^{a},p_{+},x_{+},1)= (i/{\sqrt{2}})({\psi}_{--}-(1/2){\psi}_{++})
\nonumber \\
{\phi}(x^{a},p_{+},x_{+},2)= (i/{\sqrt{2}})({\psi}_{--}+(1/2){\psi}_{++})
\nonumber \\
{\phi}(x^{a},p_{+},x_{+},12) = (1/2){\psi}_{-+}.
\label{psimaps}
\end{eqnarray}
the representation obtained in the Appendix (see (\ref{XQaction}))is identical
with the one constructed
in the produced representation in \S 2 above,
provided that the Fourier transforms of the functions on $x_-$ have support
on $p_+ \in (0,\infty)$ in conformity with the present construction.

Having established for this model the equivalence of
the physical quantisation construction with the algebraic
produced representation, we can now proceed to identify physical states
(in either picture) by computing the BRST cohomology.
The BRST-invariant states are defined by the condition
${\Omega'} |{\phi}> = 0$, with general solution $|{\phi}>= |{\psi}> +
{\Omega'}|{\chi}>$
and $|{\psi}>$ not in the range of $\Omega'$. Then using the information
of the Appendix,
the above condition
gives the following restrictions for the wave functions
${\psi}_{{\sigma}{\sigma'}}(x^{a},p_{+},{\tau})$:
\begin{eqnarray}
i{\frac{d}{dp_{+}}}{\psi}_{--}=0 \nonumber \\
H{\psi}_{--}=i{\frac{d}{d{\tau}}}{\psi}_{--}=0, \nonumber \\
(i{\frac{d}{dp_{+}}})(1/2){\psi}_{-+}
+i{\frac{d}{d{\tau}}}{\psi}_{+-}=0
\label{restrns}
\end{eqnarray}
where the Schr\"{o}dinger equation has been used for the last two
expressions. At the algebraic level the above restrictions
arise by demanding the vanishing of
$(L_{-1}+L_{-2}){\psi}_{+-}$, $(L_{-1}+L_{-2}){\psi}_{-+}$, and
$(L_{-1}+L_{-2}){\psi}_{++}$, leading to (\ref{restrns}{\it
a}-\ref{restrns}{\it c}),
respectively, while the condition
$(L_{-1}+L_{-2}){\psi}_{--}=0$
is identically satisfied. Thus the wavefunction
${\psi}_{--}$ of ghost number -1 is BRST-invariant, by
(\ref{restrns}{\it a},\ref{restrns}{\it b}) is
independent of ${\tau}=x_{+}$ and $p_{+}$,  and by (\ref{restrns}{\it b})
and (\ref{bigschroed}) satisfies the
Klein-Gordon equation. In conclusion, we see
${\psi}_{--}$ and any BRST-equivalent states of ghost number -1 are in direct
correspondence with the physical states.

These results have equivalents at the level of the produced algebra
representation via
(\ref{psimaps}). The BRST invariance conditions
become conditions for the vanishing of $(L_{-1}+L_{-2})$.
Transforming from  $x_{-}$ to $p_{+}$, and using the irreduciblity
requirement for the multiplet (with $\lambda = m^2$ in (\ref{Caseigenvalue})),
we see that
the vanishing of $(L_{-1}+L_{-2})$ on ${\phi}(x)$ will lead to
(\ref{restrns}{\it a}),
on ${\phi}(x,\alpha)$ will both lead to (\ref{restrns}{\it c}),
and on ${\phi}(x,12)$ will lead to (\ref{restrns}{\it b}).
Again, the physical states are identified with
$(-i/2){\sqrt{2}}({\phi}(x^{\mu},0,0,1)+{\phi}(x^{\mu},0,0,2))$.
Finally note that the requirement that the $iosp(d,2/2)$ states should
satisfy the Schr\"{o}dinger equation is identical with the demand that
the covariant massive scalar $iosp(d,2/2)$ multiplet should be
irreducible. That is, relation (\ref{Caseigenvalue}) should be satisfied, and
we
easily see that the effective Hamiltonian should have the
form $P_-=H=-(1/2)P^{-1}_{+}(P_{{\mu}}P^{{\mu}}-m^2+Q_{\alpha}Q^{\alpha})$.

\section{Conclusions}

In this paper we have considered in detail the canonical BFV-BRST quantisation
of the
scalar relativistic particle and its relationship to the extended quantisation
supersymmetry superalgebra $iosp(d,2/2)$. In particular, a certain type of
covariant scalar produced module of the latter is identified
with the extended state space of the particle quantisation in the usual
wavefunction
and $b,c$ algebra constructions.

Features of our approach have been the consistent treatment of the quantisation
problem
for the Lagrange multiplier on the half-line ($p_+ \equiv \lambda^{-1} > 0$ in
our
notation ) which is necessary for the identification of the $iosp(d,2/2)$
covariance (see comments below).
Although the emergence of the extended $iosp(d,2/2)$ algebra may seem
fortuitous in this particle quantisation example (\S 3), the equivalence with
the canonical
produced algebra construction (\S 2) suggests that the phenomenon is quite
universal. Thus it might be expected that the BFV-BRST quantisation
using a broad class of gauge fixing fermions corresponding to
admissible gauge fixings (see below) of the general
type\cite{govaertsbk} $\dot{\lambda}=F(\lambda)$ would also
admit the extended supersymmetry. With regard to the identification with the
produced representation,
it must be noted that the natural inner product (\S 2) is supplanted by a
pointwise inner product (\S 3) which in principle is proper-time dependent.
Of course, for states obeying Schr\"{o}dinger's equation,
this inner product is necessarily proper time {\it independent}.

The $iosp(d,2/2)$ representation (\S 3)
has been explicitly shown to be built in terms of only $d\,+\,1$ canonically
conjugate pairs of bosonic variables (together with the extended fermionic
modes),
with one momentum component, $p_-$, identified with the Hamiltonian $H$, and
its `conjugate' variable $x_+$ set equal to the proper time $\tau$.
At the $(d,2/2)$-dimensional level the realisation is
analogous to a reduced phase space or Hamiltonian reduction approach, with
constraints
solved explicitly in terms of an independent set of variables;
related constructions have also been proposed abstractly
for `covariant' quantisation algebras\cite{siegel}.

In the present work
no direct appeal is made to superfield constructions. Although in this case the
representation found can in fact be shown to be
identical to a superfield version\cite{casalbuoni2}, our approach is more
general
and is still possible for cases where superfield considerations are
inappropriate or not available.
Indeed, the general theory of produced representations as exemplified here,
provides\cite{hartley,hartleyJFC} a formal link between abstract
representation theory and more heuristic superfield methods.

Since the $iosp(d,2/2)$ covariance is established
at the level of the state space, we have not entered into considerations of
the path integral representation of the canonical action and generating
function\cite{neveuwest,casalbuoni1}. Nevertheless, for the present case the
evolution kernel can in principle be evaluated directly. The derived causal
scalar particle Green's function would then establish the connection with the
second-quantised
theory. The choice $p_+ \equiv \lambda^{-1} > 0$
corresponds, with the gauge class used, to an admissible
section\cite{govaertsbk} of the
space of gauge orbits, including the global modular transformation (in this
case
an orientation-reversing diffeomorphism, which together with the identity forms
a
$\IZ_2$ group). The quantisation is thus carried out for
the unoriented scalar particle; the opposite sign would correspond to the
unoriented scalar antiparticle\cite{govaertsbk}, and indeed the usual extension
whereby
$\phi(-p_+) \sim \phi^*(p_+)$ is consistent with this $PCT$
transformation\cite{halpern}.

It is a striking fact that both for the massless and massive particle, the
extended
quantisation symmetry involves {\it massless} representations at the $(d,2/2)$
level,
since the identification of the (inverse) Lagrange multiplier with $p_+$, and
of the
Hamiltonian with $p_-$ is perfect for the interpretation (on physical states)
of the Schr\"odinger equation
$H=-(1/2)p^{-1}_{+}(p_{\mu}p^{\mu}+Q_{\alpha}Q^{\alpha}-m^2)$  as
the vanishing of the quadratic Casimir, in light-cone coordinates for the 2
extra bosonic
directions.
The `dimensional reduction' from $(d,2/2)$ to $d$ dimensions appears
here in the analysis of physical states directly via the wavefunctions'
independence
of $x_\pm$,
rather than through a Parisi-Sourlas\cite{parisi} cancellation mechanism,
although this has been established abstractly for Greens functions in
the case of irreducible {\it induced}
representations of $iosp(d,2/2)$ by Cornwell and
Hartley\cite{hartley,hartleyJFC}.
Similar reductions have been discussed in the context of
loop integrals in quantum field theory\cite{mcclain,ricotta,ishizuka}.

Future work\cite{pdjit} in the programme initiated here will extend the
algebraic analysis to
other first quantised systems such as the spinning particle and superparticle,
as well as to gauge field theories such as Yang-Mills-Shaw. General questions
will be to
confirm the covariance of the canonical approach and ghost
systems\cite{kugoojima} with respect to an
extended orthosymplectic spacetime symmetry, particularly with regard to issues
of
modular invariance and the relation of Teichm\"{u}ller space to the appropriate
induced or produced representation theory. At this level should also emerge the
reasons for the use in the literature of $(d/2)$- as opposed to $(d,
2/2)$-dimensional
superfield formalisms for covariant quantisation and discussions of
renormalisation\cite{jarvisetal, joglekar}, and the connection with geometrical
approaches based on
coset space dimensional reduction\cite{pdj}.
Finally, the algebraic structure of quantisation using BRST symmetry is
extremely
rich and flexible, as has been demonstrated by investigations of alternative
schemes
in the context of internal symmetry\cite{bowick} and of cohomological
approaches\cite{vanholten2}. It can be expected that the study of extended
quantisation symmetries along the lines advocated here may lead to consistent
ways of implementing
covariant quantisation in systems such as string field theories where the gauge
algebra
presents technical difficulties. In any case, it is reasonable to assert
that a `Wigner' type classification of admissible `gauge multiplets'
may evolve from this viewpoint.

\subsection*{Acknowledgements}
The authors are grateful to A. J. Bracken, R. Delbourgo, D. Kreimer,
D. McAnally and J. Govaerts for helpful hints and comments which
facilitated the development of the present work.
This work was supported under grant A69332249 from the Australian Research
Council.

\section*{Appendix}
\setcounter{equation}{0}
\renewcommand{\theequation}{A.\arabic{equation}}

The action of the transformed operators $X_{\alpha}, Q_{\alpha},{\hat{\pi}}$
on the fundamental kets $|x^{\mu}, p_{+}, {\sigma}{\sigma'}>$
is given by
\begin{eqnarray}
X_{{\alpha}}|x^{{\mu}}, p_{+}, -->&=& 2(-1)^{{\alpha}}X_{{\alpha}}|x^{{\mu}},
p_{+}, ++>
 = -2(i/\sqrt{2})|x^{{\mu}}, p_{+}, -+> \nonumber\\
X_{{\alpha}}|x^{{\mu}}, p_{+}, +->&=&(i/\sqrt{2})((-1)^{{\alpha}-1}|x^{{\mu}},
p_{+},-->
 + 2 |x^{{\mu}}, p_{+},++>) \nonumber\\
Q_{{\alpha}}|x^{{\mu}}, p_{+}, -->&=&2(-1)^{{\alpha}}Q_{{\alpha}}|x^{{\mu}},
p_{+}, ++>
=i/2{\sqrt{2}}|x^{{\mu}}, p_{+}, +-> \nonumber\\
Q_{{\alpha}}|x^{{\mu}}, p_{+},-+>&=&(i/2{\sqrt{2}})( 2 |x^{{\mu}}, p_{+},++>
 + (-1)^{{\alpha}-1}|x^{{\mu}}, p_{+},-->) \nonumber\\
Q_{1}Q_{2}|x^{{\mu}}, p_{+},-+>&=& (1/2)|x^{{\mu}}, p_{+},+-> \nonumber\\
{\hat{\pi}}|x^{{\mu}}, p_{+}, {\sigma}{\sigma'}> &=&
-(ip_{+}{\frac{d}{dp_{+}}} +(i/2))|x^{{\mu}}, p_{+} {\sigma}{\sigma'}>
\nonumber\\
X_{-}|x^{{\mu}}, p_{+},{\sigma}{\sigma'}>&=&-i{\frac{d}{dp_{+}}}|x^{{\mu}},
p_{+},{\sigma}{\sigma'}>
\label{XQaction}
\end{eqnarray}
The action of any element $A$, of $iosp(d,2/2)$ or of an operator corresponding
to phase space variables, on the functions
${\psi}_{{\sigma}{\sigma'}}(x^{\mu}, p_{+}, {\tau})$ is given by
\begin{eqnarray}
A{\psi}_{{\sigma}{\sigma'}}(x^{\mu}, p_{+}, {\tau}) =
-i(-1)^{(1-{\sigma}')/2}<x^{\mu},p_{+}, -{\sigma}, -
{\sigma'}|A|\psi>
\label{Aaction}
\end{eqnarray}
We can now calculate the action of the operators (\ref{physalg}) on
${\psi}_{{\sigma}{\sigma'}}(x^{\mu},p_{+},{\tau})$ using
(\ref{XQaction}-\ref{Aaction}):
\begin{eqnarray}
J_{{\mu}{\nu}} {\psi}_{{\sigma}{\sigma'}}=i(x_{{\mu}}\frac{d}{dx^{{\nu}}}
-x_{{\nu}}\frac{d}{dx^{{\mu}}}){\psi}_{{\sigma}{\sigma'}}, \quad
P_{a}{\psi}_{{\sigma}{\sigma'}} =i
\frac{d}{dx^{a}}{\psi}_{{\sigma}{\sigma'}}\nonumber\\
K_{11}{\psi}_{--}=(1/2)K_{11}{\psi}_{++}=i{\psi}_{--} -
(i/2){\psi}_{++}\nonumber\\
K_{22}{\psi}_{--}=(-1/2)K_{22}{\psi}_{++}=-i{\psi}_{--} -
(i/2){\psi}_{++}\nonumber\\
K_{12}{\psi}_{--}= (i/2){\psi}_{++}, \quad K_{12}{\psi}_{++}=
2i{\psi}_{--}\nonumber\\
K_{{\alpha}{\beta}}{\psi}_{{\pm}{\mp}}=0\nonumber\\
Q_{{\alpha}}{\psi}_{+-}=(-1)^{{\alpha}}{\frac{i}{2 \sqrt{2} }}{\psi}_{++}+
{\frac{i}{\sqrt{2}}}{\psi}_{--} \nonumber\\
Q_{1}{\psi}_{-+}=Q_{2}{\psi}_{-+}=Q_{1}Q_{2}{\psi}_{--}=
Q_{1}Q_{2}{\psi}_{++}=Q_{1}Q_{2}{\psi}_{-+}=0 \nonumber\\
Q_{{\alpha}}{\psi}_{++}=2(-1)^{{\alpha}-1}Q_{{\alpha}}{\psi}_{--}=
\frac{i}{{\sqrt{2}}}{\psi}_{-+}\nonumber\\
Q_{1}Q_{2}{\psi}_{+-} = (1/2){\psi}_{-+}\nonumber\\
L_{-1}{\psi}_{--}=-L_{-2}{\psi}_{--}=( \frac{-i}{2 \sqrt{2} })
i{\frac{d}{dp_{+}}}{\psi}_{-+}- \frac{i}{\sqrt{2}} H{\psi}_{+-} \nonumber\\
L_{-1}{\psi}_{+-}=-i \frac{d}{dp_{+}}
(Q_{1}{\psi}_{+-}), \quad L_{-2}{\psi}_{+-}=-i{\frac{d}{dp_{+}}}
(Q_{2}{\psi}_{+-})\nonumber\\
L_{-1}{\psi}_{-+}=2H(Q_{1}{\psi}_{+-}),
\quad L_{-2}{\psi}_{-+}=2H(Q_{2}{\psi}_{+-})\nonumber\\
L_{-1}{\psi}_{++}=L_{-2}{\psi}_{++}=( \frac{-i}{{\sqrt{2}}})
i \frac{d}{dp_{+}} {\psi}_{-+}-2 \frac{i}{ \sqrt{2} }H{\psi}_{+-}\nonumber \\
L_{{\mu}{\alpha}}{\psi}_{{\sigma}{\sigma'}}
=-x_{{\mu}}Q_{\alpha}{\psi}_{{\sigma}{\sigma'}} +
i{\frac{d}{dx^{{\mu}}}}X_{\alpha}{\psi}_{{\sigma}{\sigma'}}\nonumber\\
J_{-{\mu}} {\psi}_{{\sigma}{\sigma'}}=
i{\frac{d}{dx^{{\mu}}}}(i{\frac{d}{dp_{+}}}){\psi}_{{\sigma}{\sigma'}}
 + x_{{\mu}}H{\psi}_{{\sigma}{\sigma'}},\nonumber\\
J_{+-} {\psi}_{{\sigma}{\sigma'}}=x_{+}H{\psi}_{{\sigma}{\sigma'}}
+i{\frac{d}{dp_{+}}}
(p_{+}{\psi'}_{{\sigma}{\sigma'}})
\label{ospmodelaction}
\end{eqnarray}
{}From the Schr\"{o}dinger equation
\begin{eqnarray}
i{\frac{d}{d{\tau}}}|{\psi}_{{\sigma}{\sigma'}}> =
 H|{\psi}_{{\sigma}{\sigma'}}>
\label{schroed}
\end{eqnarray}
and (\ref{XQaction}-\ref{Aaction}) we also have
\begin{eqnarray}
i{\frac{d}{d{\tau}}}{\psi}_{--} = H{\psi}_{--} =
-{\frac{1}{2}}p^{-1}_{+}(P^{2} - m^{2}){\psi}_{--}\nonumber\\
i{\frac{d}{d{\tau}}}{\psi}_{++} = H{\psi}_{++} =
-{\frac{1}{2}}p^{-1}_{+}(P^{2} - m^{2}){\psi}_{++}\nonumber\\
i{\frac{d}{d{\tau}}}{\psi}_{+-}=H{\psi}_{+-}=
-{\frac{1}{2}}p^{-1}_{+}
(P^{2} - m^{2}){\psi}_{+-} +ip^{-1}_{+}Q_{1}Q_{2}{\psi}_{+-}\nonumber\\
i{\frac{d}{d{\tau}}}{\psi}_{-+} = H{\psi}_{-+} =
-{\frac{1}{2}}p^{-1}_{+}(P^{2} - m^{2}){\psi}_{-+}.
\label{bigschroed}
\end{eqnarray}

\end{document}